\documentclass[12pt]{article}
\include{epsf}
\include{psfig}
\include{epsfig}
\usepackage{epsfig}
\makeatletter

\def\lsim{\compoundrel<\over\sim}
\def\compoundrel#1\over#2{\mathpalette\compoundreL{{#1}\over{#2}}}
\def\compoundreL#1#2{\compoundREL#1#2}
\def\compoundREL#1#2\over#3{\mathrel
         {\vcenter{\hbox{$\m@th\buildrel{#1#2}\over{#1#3}$}}}}
\makeatother

\makeatletter
\def\sss{\scriptscriptstyle}
\def\sVEV#1{\left\langle #1\right\rangle}

\newcommand{\MeV}{\mbox{\rm MeV}}
\newcommand{\GeV}{\mbox{\rm GeV}}
\newcommand{\eV}{\mbox{\rm eV}}
\def\sleq{\raisebox{-.6ex}{${\textstyle\stackrel{<}{\sim}}$}}

\begin{document}
\setcounter{page}{0}
\thispagestyle{empty}
\setlength{\parindent}{1.0em}
\begin{flushright}
GUTPA/03/12/02
\end{flushright}
\renewcommand{\thefootnote}{\fnsymbol{footnote}}
\begin{center}{\LARGE{{\bf The Problem of Mass\footnote{\it To be 
published in the Proceedings of the Euresco 
Conference on What comes beyond the Standard Model? Symmetries beyond 
the Standard Model, Portoroz, Slovenia, 12 - 17 July 2003.} }}}
\end{center}
\begin{center}{\large{C. D. Froggatt}
\\}
\end{center}
\renewcommand{\thefootnote}{\arabic{footnote}}
\begin{center}{{\it Department of Physics and Astronomy}\\{\it
University of Glasgow, Glasgow G12 8QQ, Scotland}}\end{center}

\setcounter{footnote}{0}

\begin{abstract}
The quark-lepton mass problem and the ideas of mass protection are
reviewed. The Multiple Point Principle is introduced and used
within the Standard Model to predict the top quark and Higgs
particle masses. We discuss the lightest family mass generation
model, in which all the quark mixing angles are successfully
expressed in terms of simple expressions involving quark mass
ratios. The chiral flavour symmetry of the family replicated gauge
group model is shown to provide the mass protection needed to
generate the hierarchical structure of the quark-lepton mass
matrices.
\end{abstract}
\thispagestyle{empty}

\newpage

\section{Introduction}
\label{introduction}

The most important unresolved problem in particle physics is the
understanding of flavour and the fermion mass spectrum. The observed
values of the fermion masses and mixing angles constitute the bulk of
the Standard Model (SM) parameters and provide our main experimental clues
to the underlying flavour dynamics. In particular the non-vanishing
neutrino masses and mixings provide direct evidence for physics beyond
the SM.

The charged lepton masses can be directly measured and correspond to
the poles in their propagators:
\begin{equation}
M_e = 0.511 \ \makebox{MeV} \qquad M_{\mu} = 106 \ \makebox{MeV}
\qquad M_{\tau} = 1.78 \ \makebox{GeV}
\end{equation}
However the quark masses have to be extracted from the properties of
hadrons and are usually quoted as running masses $m_q(\mu)$ evaluated at
some renormalisation scale $\mu$, which are related to the  propagator
pole masses $M_q$ by
\begin{equation}
M_q = m_q(\mu=m_q)\left[ 1+ \frac{4}{3}\alpha_3(m_q) \right]
\end{equation}
to leading order in QCD. The light $u$, $d$ and $s$ quark masses
are usually normalised to the scale $\mu=1$ GeV (or $\mu=2$ GeV
for lattice measurements) and to the quark mass itself for the
heavy $c$, $b$ and $t$ quarks. They are typically given
\cite{pdg} as follows\footnote{Note that the top quark mass,
$M_t = 174 \pm 5$ GeV, measured at FermiLab is interpreted as
the pole mass.}:
\begin{eqnarray}
m_u(1 \ \makebox{GeV})  = 4.5 \pm 1 \ \makebox{MeV} &\qquad &
m_d(1 \ \makebox{GeV})  = 8 \pm 2 \ \makebox{MeV} \nonumber \\
m_c(m_c)  =  1.25 \pm 0.15 \ \makebox{GeV} &\qquad &
m_s(1 \ \makebox{GeV}) =  150 \pm 50 \ \makebox{MeV} \nonumber \\
m_t(m_t)  =  166 \pm 5 \ \makebox{GeV} &\qquad &
m_b(m_b)  =  4.25 \pm 0.15 \ \makebox{GeV}
\end{eqnarray}
However we only have an upper limit on the neutrino masses of
$m_{\nu_i} \lsim 1$ eV from tritium beta decay and from cosmology,
and measurements of the neutrino mass squared differences:
\begin{equation}
 \Delta m_{21}^2 \sim 5 \times 10^{-5} \makebox{eV}^2 \qquad
 \Delta m_{32}^2 \sim 3 \times 10^{-3} \makebox{eV}^2
 \label{dm2}
\end{equation}
from solar and atmospheric neutrino oscillation data \cite{garcia}.

The magnitudes of the quark mixing matrix $V_{CKM}$ are well measured
\begin{equation}
 |V_{CKM}| = \pmatrix
{0.9734 \pm 0.0008 & 0.2196 \pm 0.0020 & 0.0036 \pm 0.0007 \cr
0.224 \pm 0.016   & 0.996 \pm 0.013   & 0.0412 \pm 0.002 \cr
0.0077 \pm 0.0014 & 0.0397 \pm 0.0033 & 0.9992 \pm 0.0002 \cr}
\end{equation}
and a CP violating phase of order unity:
\begin{equation}
\sin^2\delta_{CP} \sim 1
\end{equation}
can reproduce all the CP violation data. Neutrino
oscillation data constrain the magnitudes of the lepton
mixing matrix elements to lie in the following $3\sigma$ ranges
\cite{garcia}:
\begin{equation}
 |U_{MNS}| = \pmatrix
 {0.73-0.89 & 0.45-0.66 & <0.24     \cr
 0.23-0.66  & 0.24-0.75 & 0.52-0.87 \cr
 0.06-0.57  & 0.40-0.82 & 0.48-0.85 \cr}
 \label{mns}
\end{equation}
Due to the Majorana nature of the neutrino mass matrix, there are
three unknown CP violating phases $\delta$, $\alpha_1$ and
$\alpha_2$ in this case \cite{garcia}.

The charged fermion masses range over five orders of magnitude,
whereas there seems to be a relatively mild neutrino mass hierarchy.
The absolute neutrino mass scale ($m_{\nu} < 1$
eV) suggests a new physics mass scale -- the
so-called see-saw scale $\Lambda_{seesaw} \sim 10^{15}$ GeV.
The quark mixing matrix $V_{CKM}$ is also hierarchical, with small
off-diagonal elements. However the elements of $U_{MNS}$ are all
of the same order of magnitude except for $|U_{e3}|<0.24$,
corresponding to two leptonic mixing angles being close to
maximal ($\theta_{atmospheric} \simeq \pi/4$ and $\theta_{solar}
\simeq \pi/6$).

We introduce the mechanism of mass protection by approximately
conserved chiral charges in section \ref{protection}. The top
quark mass is the dominant term in the SM fermion mass matrix, so
it is likely that its value will be understood dynamically before
those of the other fermions. In section \ref{top} we discuss the
connection between the top quark and Higgs masses and how they can
be determined from the so-called Multiple Point Principle. We
present the lightest family mass generation model in section
\ref{ansatze}, which provides an ansatz for the texture of fermion
mass matrices and expresses all the quark mixing angles
successfully in terms of simple expressions involving quark mass
ratios. The family replicated gauge group model is presented in
section \ref{origin}, as an example of a model whose gauge group
naturally provides the mass protecting quantum numbers needed to
generate the required texture for the fermion mass matrices.
Finally we present a brief conclusion in section \ref{conclusion}.

\section{Fermion Mass and Mass Protection}
\label{protection}

A fermion mass term
\begin{equation}
{\cal{L}}_{mass} =  -m \overline{\psi}_L \psi_R + h. c.
\end{equation}
couples together a left-handed Weyl field $\psi_L$ and a
right-handed Weyl field $\psi_R$, which then satisfy the Dirac
equation
\begin{equation}
 i\gamma^{\mu} \partial_{\mu} \psi_L = m \psi_R
\end{equation}
If the two Weyl fields are not charge conjugates $\psi_L \neq
(\psi_R)^c$ we have a Dirac mass term and the two fields $\psi_L$
and $\psi_R$ together correspond to a Dirac spinor. However if the
two Weyl fields are charge conjugates $\psi_L = (\psi_R)^c$ we
have a Majorana mass term and the corresponding four component
Majorana spinor has only two degrees of freedom. Particles
carrying an exactly conserved charge, like the electron carries
electric charge, must be distinct from their anti-particles and
can only have Dirac masses with $\psi_L$ and $\psi_R$ having equal
conserved charges. However a neutrino could be a Majorana
particle.

If $\psi_L$ and $\psi_R$ have different quantum numbers,
i.e.~belong to inequivalent representations of a symmetry group
$G$ ($G$ is then called a chiral symmetry), a Dirac mass term is
forbidden in the limit of an exact $G$ symmetry and they represent
two massless Weyl particles. Thus the $G$ symmetry ``protects'' the
fermion from gaining a mass. Such a fermion can only gain a mass
when $G$ is spontaneously broken.

The left-handed and right-handed top quark, $t_L$ and $t_R$, carry
unequal Standard Model $SU(2) \times U(1)$ gauge charges
$\vec{Q}$:
\begin{equation}
 \vec{Q}_L \neq \vec{Q}_R \qquad \mathrm{(Chiral\  charges)}
\end{equation}
Hence electroweak gauge invariance protects the quarks and leptons
from gaining a fundamental mass term ($\overline{t}_L t_R$ is not
gauge invariant). This {\em mass protection} mechanism is of
course broken by the Higgs effect, when the vacuum expectation
value of the Weinberg-Salam Higgs field
\begin{equation}
 <\phi_{WS}> = \sqrt{2} v = 246 \ GeV
\end{equation}
breaks the gauge symmetry and the SM gauge invariant Yukawa
couplings $\frac{y_i}{\sqrt{2}}$ generate the running quark masses
$m_i = y_i v  =  174 \, y_i \ \makebox{GeV}$. In this way a top
quark mass of the same order of magnitude as the SM Higgs field
vacuum expectation value (VEV) is naturally generated (with $y_t$
unsuppressed). Thus the Higgs mechanism explains why the top quark
mass is suppressed, relative to the fundamental (Planck, GUT...)
mass scale of the physics beyond the SM, down to the scale of
electroweak gauge symmetry breaking. However the further
suppression of the other quark-lepton masses ($y_b$, $y_c$, $y_s$,
$y_u$, $y_d$ $\ll$ 1) remains a mystery, which it is natural to
attribute to mass protection by other approximately conserved
chiral gauge charges beyond the SM, as discussed in section
\ref{origin} for the family replicated gauge group model.

Fermions which are vector-like under the SM gauge group
($\vec{Q}_L = \vec{Q}_R$) are not mass protected and are expected
to have a large mass associated with new (grand unified,
string,..) physics. The Higgs particle, being a scalar, is not
mass protected and {\em a priori} would also be expected to have a
large mass; this is the well-known gauge hierarchy problem
discussed at Portoroz by Holger Nielsen \cite{holger}.

\section{Top Quark and Higgs Masses from the Multiple Point Principle}
\label{top}

\begin{figure}
\leavevmode 
\centerline{
\epsfig{file=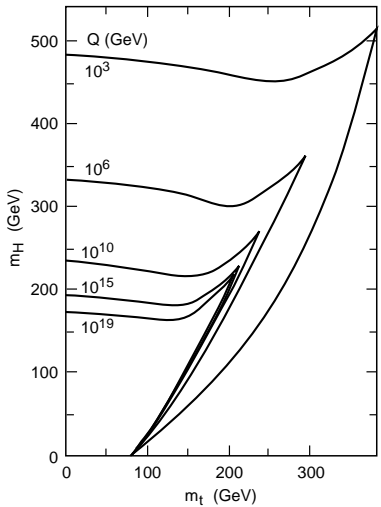,width=5.0cm}} 
\caption{SM bounds in the ($m_t$,$m_H$) plane for  various values
of $\Lambda = Q$, the scale at which new physics enters.}
\label{fig:Maiani}
\end{figure}
It is well-known \cite{maiani} that the self-consistency of the
pure SM up to some physical cut-off scale $\Lambda$ imposes
constraints on the top quark and Higgs boson masses. The first
constraint is the so-called triviality bound: the running Higgs
coupling constant $\lambda(\mu)$ should not develop a Landau pole
for $\mu < \Lambda$. The second is the vacuum stability bound: the
running Higgs coupling constant $\lambda(\mu)$ should not become
negative leading to the instability of the usual SM vacuum. These
bounds are illustrated \cite{lindner} in Fig. \ref{fig:Maiani},
where the combined triviality and vacuum stability bounds for the
SM are shown for different values of the high energy cut-off
$\Lambda$. The allowed region is the area around the origin
bounded by the co-ordinate axes and the solid curve labelled by
the appropriate value of $\Lambda$.  The upper part of each curve
corresponds to the triviality bound. The lower part of each curve
coincides with the vacuum stability bound and the point in the top
right hand corner, where it meets the triviality bound curve, is
the infra-red quasi-fixed point for that value of $\Lambda$. Here
the vacuum stability curve, for a large cut-off of order the
Planck scale $\Lambda_{Planck} \simeq 10^{19}$ GeV, is important
for the discussion of the values of the top quark and Higgs boson
masses predicted from the Multiple Point Principle.

\begin{figure}
\leavevmode \vspace{-0.5cm}
\centerline{\epsfig{file=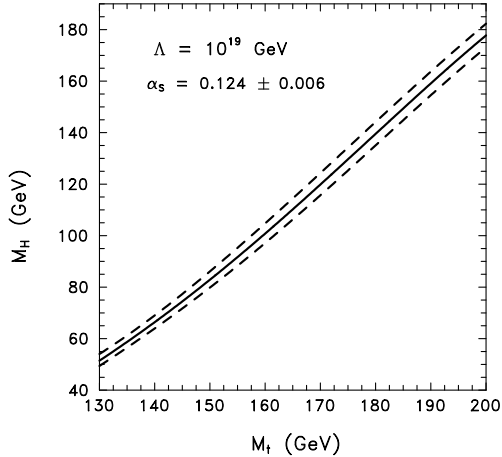,width=7cm,%
bbllx=95pt,bblly=130pt,bburx=510pt,bbury=555pt,%
angle=270,clip=} } \caption{SM vacuum stability curve for $\Lambda
= 10^{19}$ GeV and $\alpha_s = 0.124$ (solid line), $\alpha_s =
0.118$ (upper dashed line), $\alpha_s = 0.130$ (lower dashed
line).} \label{fig:vacstab}
\end{figure}
According to the Multiple Point Principle (MPP), Nature chooses
coupling constant values such that a number of vacuum states have
the same energy density (cosmological constant). This fine-tuning
of the coupling constants is similar to that of temperature for a
mixture of co-existing phases such as ice and water. We have
previously argued \cite{glasgowbrioni} that baby-universe like
theories \cite{baby}, having a mild breaking of locality and
causality, may contain the underlying physical explanation of the
MPP, but it really has the status of a postulated new principle.
Here we apply it to the pure Standard Model \cite{fn2}, which we
assume valid up close to $\Lambda_{Planck}$. So we shall postulate
that the effective potential $V_{eff}(\phi)$ for the SM Higgs
field $\phi$ should have a second minimum, at $<\phi> =
\phi_{vac\; 2}$, degenerate with the
well-known first minimum at the electroweak scale $<\phi> =
\phi_{vac\; 1} = 246$ GeV:
\begin{equation}
V_{eff}(\phi_{vac\; 1}) = V_{eff}(\phi_{vac\; 2}) \label{eqdeg}
\end{equation}
Thus we predict that our vacuum is barely stable and we just lie
on the vacuum stability curve in the top quark, Higgs particle
(pole) mass ($M_t$, $M_H$) plane, shown \cite{casas} in Fig.
\ref{fig:vacstab} for a cut-off $\Lambda = 10^{19}$ GeV.
Furthermore we expect the second minimum to be within an order of
magnitude or so of the fundamental scale, i.e. $\phi_{vac\; 2}
\simeq \Lambda_{Planck}$. In this way, we essentially select a
particular point on the SM vacuum stability curve and hence the
MPP condition predicts precise values for $M_t$ and $M_H$.
\begin{figure}
\leavevmode \centerline{ \epsfig{file=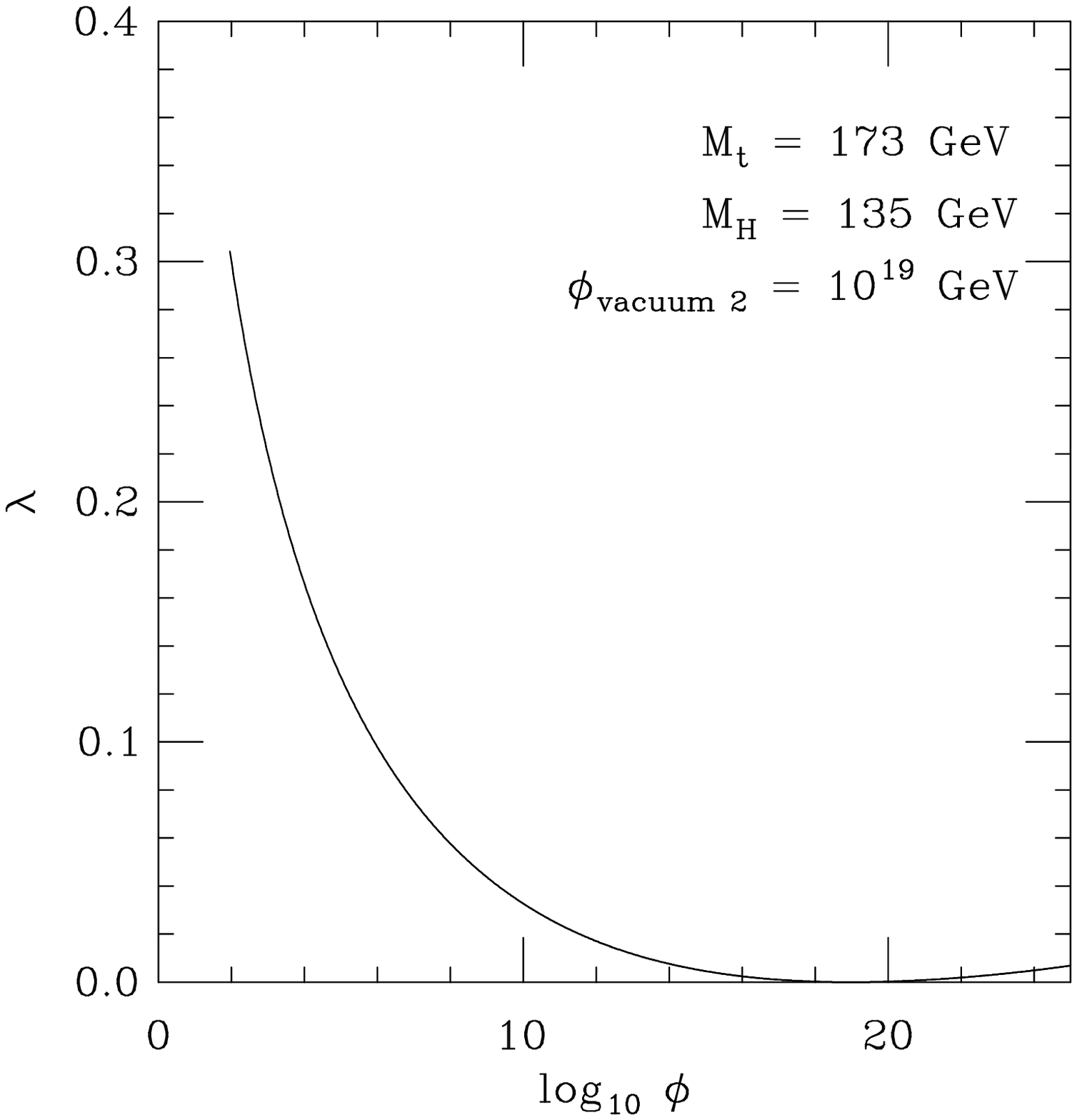,width=6.8cm}
\hspace{-0.6cm} \epsfig{file=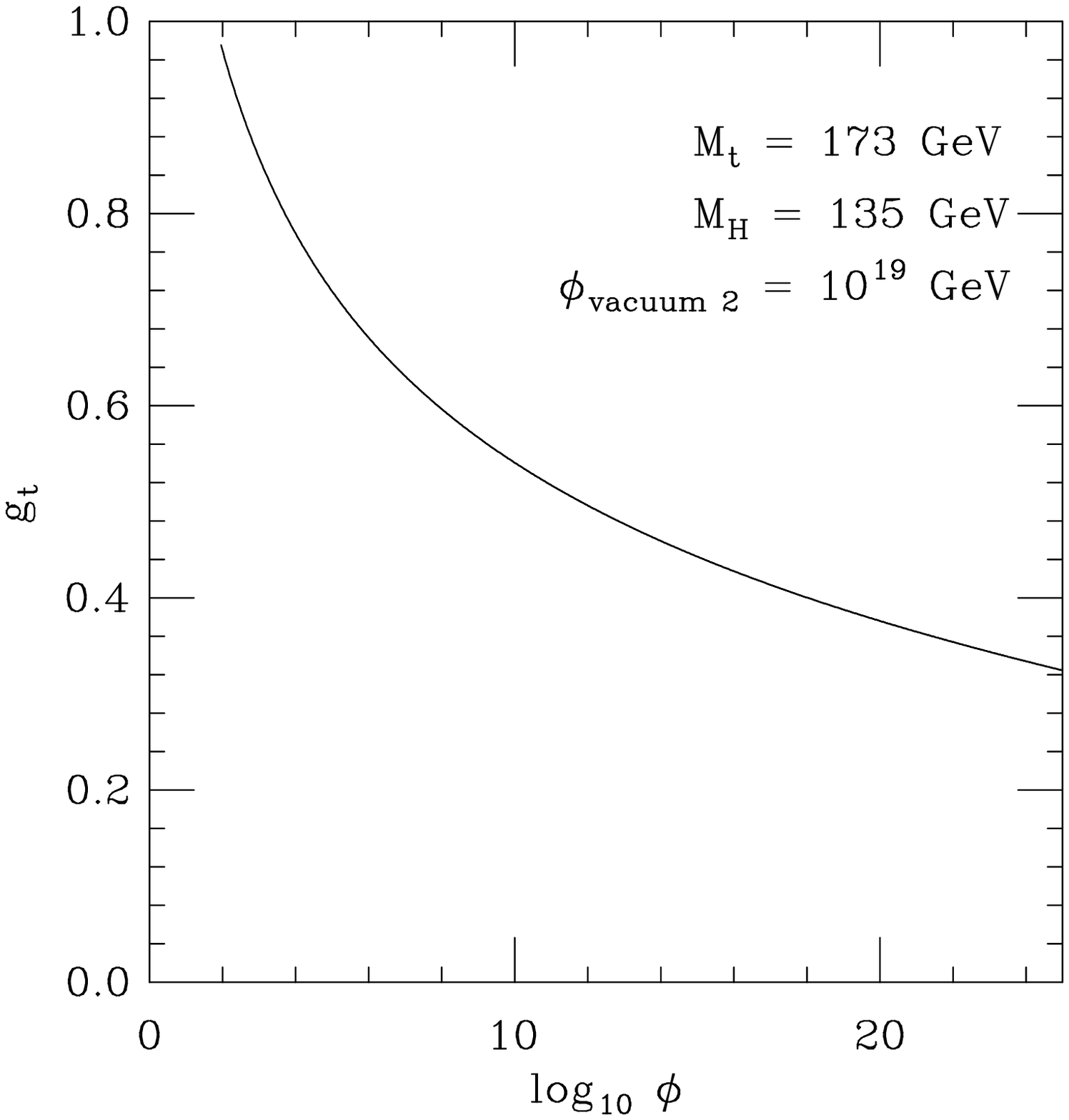,width=6.8cm} }
\vspace{-0.6cm} \caption{Plots of $\lambda$ and $g_t$ as functions
of the scale of the Higgs field $\phi$ for degenerate vacua with
the second Higgs VEV at the Planck scale $\phi_{vac\;2}=10^{19}$
GeV. We formally apply the second order SM renormalisation group
equations up to a scale of $10^{25}$ GeV.} \label{fig:lam19}
\end{figure}

For large values of the SM Higgs field $\phi >> \phi_{vac\; 1}$,
the renormalisation group improved tree level effective potential
is very well approximated by $V_{eff}(\phi) \simeq
\frac{1}{8}\lambda (\mu = |\phi | ) |\phi |^4$ and the degeneracy
condition, eq.~(\ref{eqdeg}), means that $\lambda(\phi_{vac\; 2})$
should vanish to high accuracy. The derivative of the effective
potential $V_{eff}(\phi)$ should also be zero at $\phi_{vac\; 2}$,
because it has a minimum there. Thus at the second minimum of the
effective potential the beta function $\beta_{\lambda}$ vanishes
as well:
\begin{equation}
\beta_{\lambda}(\mu = \phi_{vac\; 2}) = \lambda(\phi_{vac\; 2}) = 0
\end{equation}
which gives to leading order the relationship:
\begin{equation}
\frac{9}{4}g_2^4 + \frac{3}{2}g_2^2g_1^2 + \frac{3}{4}g_1^4 -
12g_t^4 = 0
\end{equation}
between the top quark Yukawa coupling $g_t(\mu)$ and the electroweak
gauge coupling constants $g_1(\mu)$ and $g_2(\mu)$ at the scale
$\mu = \phi_{vac\; 2} \simeq \Lambda_{Planck}$. We use the
renormalisation group equations to relate the couplings at the
Planck scale to their values at the electroweak scale. Figure
\ref{fig:lam19} shows the running coupling constants
$\lambda(\phi)$ and $g_t(\phi)$ as functions of $\log(\phi)$.
Their values at the electroweak scale give our predicted
combination of pole masses \cite{fn2}:
\begin{equation}
M_{t} = 173 \pm 5\ \mbox{GeV} \quad M_{H} = 135 \pm 9\ \mbox{GeV}
\end{equation}

We have also considered \cite{metaMPP} a slightly modified version
of MPP, according to which the two vacua are approximately
degenerate in such a way that they should both be physically
realised over comparable amounts of space-time four volume. This
modified MPP corresponds to the Higgs mass lying on the vacuum
{\em metastability} curve rather than on the vacuum stability
curve, giving a Higgs mass prediction of $122 \pm 11$ GeV. We
should presumably not really take the MPP predictions to be more
accurate than to the order of magnitude of the variation between
the metastability and stability bounds. However we definitely
predict a light Higgs mass in this range, as seems to be in
agreement with indirect estimates of the SM Higgs mass from
precision data \cite{pdg}.

This application of the MPP assumes the existence of the hierarchy
$v/\Lambda_{Planck} \sim 10^{-17}$. Recently we have speculated
\cite{itepfn} that this huge scale ratio is a consequence of the
existence of yet another vacuum in the SM, at the electroweak
scale and degenerate with the two vacua discussed above. The two
SM vacua at the electroweak scale are postulated to differ by the
condensation of an S-wave bound state formed from 6 top and 6
anti-top quarks mainly due to Higgs boson exchange forces. This
scenario is discussed in more detail in Holger Nielsen's talk
\cite{holger}.

\section{Lightest Family Mass Generation Model}
\label{ansatze}

Motivated by the famous Fritzsch ansatz \cite{fritzsch} for the two
generation quark mass matrices:
\begin{equation}
M_U =\pmatrix{0      & B\cr
          B^{\ast}   & A\cr}
\qquad
M_D =\pmatrix{0          & B^\prime\cr
          B^{\prime\ast}  & A^\prime\cr}
\label{fritzsch_ansatz}
\end{equation}
several ans\"{a}tze have been proposed for the fermion mass
matrices---for example, see
\cite{rrr} for a systematic analysis of symmetric quark mass
matrices with texture zeros at the SUSY-GUT scale.
Here I will concentrate on the lightest family mass generation
model \cite{lfm}. It successfully generalizes the well-known
formula
\begin{equation}
\left| V_{us} \right| \simeq
\left|\sqrt{\frac{m_d}{m_s}} -
e^{i\phi} \sqrt{\frac{m_u}{m_c}} \right|
\label{fritzsch1}
\end{equation}
for the Cabibbo angle derived from the above ansatz,
eq.~(\ref{fritzsch_ansatz}), to simple working formulae for all
the quark mixing angles in terms of quark mass ratios. According
to this model the flavour mixing for quarks is basically
determined by the mechanism responsible for generating the
physical masses of the up and down quarks, $m_{u}$ and $m_{d}$
respectively. So, in the chiral symmetry limit, when $m_{u}$ and
$m_{d}$ vanish, all the quark mixing angles vanish. Therefore we
are led to consider an ansatz in which the diagonal mass matrix
elements for the second and third generations are practically the
same in the gauge (unrotated) and physical bases.

The mass matrix for the down quarks ($D$ = $d$, $s$, $b$) is taken
to be hermitian with three texture zeros of the following form:
\begin{equation}
M_{D}=\pmatrix{ 0 & a_D & 0 \cr a_D^{\ast} & A_D & b_D \cr 0 &
b_D^{\ast} & B_D \cr}
\label{LFM1}
\end{equation}
where
\begin{equation}
B_{D}=m_{b}+\delta_{D} \qquad A_{D}= m_{s} + \delta_{D}^{\prime }
\qquad |\delta_D |\ll m_{s} \qquad |\delta _{D}^{\prime }|\ll m_{d}
\label{BA}
\end{equation}
It is, of course, necessary to assume some hierarchy between the
elements, which we take to be: $B_{D}\gg A_{D}\sim \left|
b_{D}\right| \gg \left| a_{D}\right| $. The zero in the $\left(
M_{D}\right) _{11}$ element corresponds to the commonly accepted
conjecture that the lightest family masses appear as a direct
result of flavour mixings. The zero in $\left( M_{D}\right) _{13}$
means that only minimal ``nearest neighbour'' interactions occur,
giving a tridiagonal matrix structure. Since the trace and
determinant of the hermitian matrix $M_{D}$ gives the sum and
product of its eigenvalues, it follows that
\begin{equation}
\delta _{D}\simeq - m_{d}  \label{del}
\end{equation}
while $\delta _{D}^{\prime }$ is vanishingly small and can be neglected
in further considerations.

It may easily be shown that equations (\ref{LFM1} - \ref{del}) are
entirely equivalent to the condition that the diagonal
elements ($A_{D}$, $B_{D}$) of $M_{D}$ are proportional
to the modulus square of the off-diagonal elements ($a_{D}$, $b_{D}$):
\begin{equation}
\frac{A_{D}}{B_{D}}=\left| \frac{a_{D}}{b_{D}}\right| ^{2}
\label{ABab}
\end{equation}
Using the conservation of the trace, determinant and sum of
principal minors of the hermitian matrix $M_{D}$ under unitary
transformations, we are led to a complete determination of the
moduli of all its elements, which can be expressed to high
accuracy as follows:
\begin{equation}
\left| M_{D} \right| = \pmatrix{ 0 & \sqrt{m_d m_s} & 0 \cr
\sqrt{m_d m_s} & m_s & \sqrt{m_d m_b} \cr 0 &
\sqrt{m_d m_b} & m_b - m_d \cr}  \label{LFM1A}
\end{equation}

The mass matrix for the up quarks is taken to be of the following
hermitian form:
\begin{equation}
M_{U}=\pmatrix{ 0 & 0 & c_U \cr 0 & A_U & 0 \cr c_U^{\ast} & 0 & B_ U \cr}
\label{LFM2}
\end{equation}
The moduli of all the elements of $M_{U}$ can also be readily
determined in terms of the physical masses as follows:
\begin{equation}
\left| M_{U} \right| = \pmatrix{ 0 & 0 & \sqrt{m_u m_t} \cr
0 & m_c & 0 \cr \sqrt{m_u m_t} & 0 & m_t - m_u \cr}
\label{LFM2A}
\end{equation}

The CKM quark mixing matrix elements can now be readily calculated
by diagonalising the mass matrices $M_D$ and $M_U$. They are
given  in terms of quark mass ratios as follows:
\begin{eqnarray}
\left| V_{us}\right| = \sqrt{\frac{m_{d}}{m_{s}}} = 0.222 \pm 0.004
\qquad \left| V_{us}\right|_{exp} = 0.221 \pm 0.003 \\
\left|V_{cb}\right| = \sqrt{\frac{m_{d}}{m_{b}}} = 0.038 \pm 0.004
\qquad \left|V_{cb}\right|_{exp} = 0.039 \pm 0.003 \\
\left|V_{ub}\right| = \sqrt{\frac{m_{u}}{m_{t}}} = 0.0036 \pm 0.0006
\qquad \left|V_{ub}\right|_{exp} = 0.0036 \pm 0.0006  \\
\left|V_{td}\right| = \left|V_{us}V_{cb}-V_{ub}\right|
= 0.009 \pm 0.002
\qquad \left|V_{ub}\right|_{exp} = 0.0077 \pm 0.0014
\label{angles}
\end{eqnarray}
As can be seen, they are in impressive agreement with the experimental
values. The MNS lepton mixing matrix can also be fitted, if the texture
of eq.~(\ref{LFM1}) is extended to the Dirac and Majorana right-handed
neutrino mass matrices \cite{matsuda}.

The proportionality condition, eq.~(\ref{ABab}), is not so easy to
generate from an underlying symmetry beyond the Standard Model,
but it is possible to realise it in a local chiral $SU(3)$ family
symmetry\footnote{See ref.~\cite{SU3ross} for a local chiral
SU(3) family model with an alternative texture.} model \cite{SU3}.

\section{Family Replicated Gauge Group Model}
\label{origin}

As pointed out in section \ref{protection}, a natural explanation
of the charged fermion mass hierarchy would be mass protection due
to the existence of some approximately conserved chiral charges
beyond the SM. An attractive possibility is that these chiral
charges arise as a natural feature of the gauge symmetry group of
the fundamental theory beyond the SM. This is the case in the
family replicated gauge group model (also called the anti-grand
unification model) \cite{smg3m,fnt}. The new chiral charges
provide selection rules forbidding the transitions between the
various left-handed and right-handed quark-lepton states, except
for the top quark. In order to generate mass terms for the other
fermion states, we have to introduce new Higgs fields, which break
the symmetry group $G$ of the fundamental theory down to the SM
group. We also need suitable intermediate fermion states to
mediate the forbidden transitions, which we take to be vector-like
Dirac fermions with a mass of order the fundamental scale $M_F$ of
the theory. In this way effective SM Yukawa coupling constants are
generated \cite{fn1}, which are suppressed by the appropriate
product of Higgs field VEVs measured in units of $M_F$. We assume
that all the couplings in the fundamental theory are unsuppressed,
i.e.~they are all naturally of order unity.

The family replicated gauge group model is based on a non-simple
non-supersymmetric extension of the SM with three copies of the SM
gauge group---one for each family or generation. With the
inclusion of three right-handed neutrinos, the gauge group becomes
$G = (SMG \times U(1)_{B-L})^3$, where the three copies of the SM
gauge group are supplemented by an abelian $(B-L)$ (= baryon
number minus lepton number) gauge group for each
family\footnote{The family replicated gauge groups $(SO(10))^3$
and $(E_6)^3$ have recently been considered by Ling and Ramond
\cite{ling}.}. The gauge group $G$ is the largest anomaly free
group, transforming the known 45 Weyl fermions plus the three
right-handed neutrinos into each other unitarily, which does {\em
not} unify the irreducible representations under the SM gauge
group. It is supposed to be effective at energies near to the
Planck scale, $M_F = \Lambda_{Planck}$, where the $i$'th
proto-family couples to just the $i$'th group factor $SMG_i\times
U(1)_{B_i-L_i}$. The gauge group $G$ is broken down by four Higgs
fields $W$, $T$, $\rho$ and $\omega$, having VEVs about one order
of magnitude lower than the Planck scale, to its diagonal
subgroup:
\begin{equation}
(SMG \times U(1)_{B-L})^3 \rightarrow SMG\times U(1)_{B-L}
\end{equation}
The diagonal $U(1)_{B-L}$ is broken down at the see-saw scale, by
another Higgs field $\phi_{SS}$, and the diagonal $SMG$ is broken
down to $SU(3) \times U(1)_{em}$ by the Weinberg-Salam Higgs field
$\phi_{WS}$.

\begin{table}[!th]
\caption{All $U(1)$ quantum charges of the Higgs fields in the
$(SMG \times U(1)_{B-L})^3$ model.} \vspace{3mm} \label{qc}
\begin{center}
\begin{tabular}{|c||c|c|c|c|c|c|} \hline
& $y_1/2$& $y_2/2$ & $y_3/2$ & $(B-L)_1$ & $(B-L)_2$ & $(B-L)_3$
\\ \hline\hline
$\omega$ & $\frac{1}{6}$ & $-\frac{1}{6}$ & $0$ & $0$ & $0$ & $0$\\
$\rho$ & $0$ & $0$ & $0$ & $-\frac{1}{3}$ & $\frac{1}{3}$ & $0$\\
$W$ & $0$ & $-\frac{1}{2}$ & $\frac{1}{2}$ & $0$ & $-\frac{1}{3}$
& $\frac{1}{3}$ \\
$T$ & $0$ & $-\frac{1}{6}$ & $\frac{1}{6}$ & $0$ & $0$ & $0$\\
$\phi_{\sss WS}$ & $0$ & $\frac{2}{3}$ & $-\frac{1}{6}$ & $0$
& $\frac{1}{3}$ & $-\frac{1}{3}$ \\
$\phi_{\sss SS}$ & $0$ & $1$ & $-1$ & $0$ & $2$ & $0$ \\
\hline
\end{tabular}
\end{center}
\end{table}
The $(SMG \times U(1)_{B-L})^3$ gauge quantum numbers of the
quarks and leptons are uniquely determined by the structure of the
model and they include 6 chiral abelian charges---the weak
hypercharge $y_i/2$ and $(B-L)_i$ quantum number for each of the
three families, $i=1,2,3$. With the choice of the abelian charges
in Table \ref{qc} for the Higgs fields, it is possible to generate
a good order of magnitude fit to the SM fermion masses, with VEVs
of order $M_F/10$. In this fit, we do not attempt to guess the
spectrum of superheavy fermions at the Planck scale, but simply
assume a sufficiently rich spectrum to mediate all of the symmetry
breaking transitions in the various mass matrix elements. Then,
using the quantum numbers of Table \ref{qc}, the suppression
factors are readily calculated as products of Higgs field VEVs
measured in Planck units for all the fermion Dirac mass matrix
elements\footnote{For clarity we distinguish between Higgs fields
and their hermitian conjugates.}, giving for example:
\begin{eqnarray}
M_{\sss U} \simeq \frac{\sVEV{(\phi_{\sss\rm WS})^\dagger}}
{\sqrt{2}}\hspace{-0.1cm} \left(\!\begin{array}{ccc}
        (\omega^\dagger)^3 W^\dagger T^2
        & \omega \rho^\dagger W^\dagger T^2
        & \omega \rho^\dagger (W^\dagger)^2 T\\
        (\omega^\dagger)^4 \rho W^\dagger T^2
        &  W^\dagger T^2
        & (W^\dagger)^2 T\\
        (\omega^\dagger)^4 \rho
        & 1
        & W^\dagger T^\dagger
\end{array} \!\right)\label{M_U}
\end{eqnarray}
for the up quarks. Similarly the right-handed neutrino Majorana
mass matrix is of order:
\begin{eqnarray}
M_R \simeq \sVEV{\phi_{\sss\rm SS}}\hspace{-0.1cm} \left
(\hspace{-0.1 cm}\begin{array}{ccc} (\rho^\dagger)^6 T^6 &
(\rho^\dagger)^3 T^6
& (\rho^\dagger)^3 W^3 (T^\dagger)^3 \\
(\rho^\dagger)^3 T^6
& T^6 & W^3 (T^\dagger)^3 \\
(\rho^\dagger)^3 W^3 (T^\dagger)^3 & W^3 (T^\dagger)^3 & W^6
(T^\dagger)^{12}
\end{array} \hspace{-0.1 cm}\right ) \label{Mmajo}
\end{eqnarray}
and the effective light neutrino mass matrix can be calculated
from the Dirac neutrino mass matrix $M_N$ and $M_R$ using the
see-saw formula \cite{seesaw}:
\begin{equation}
M_{\nu} = M_N M_R^{-1} M_N^T
\end{equation}
In this way we obtain a good 5 parameter fit to the orders of
magnitude of all the quark-lepton masses and mixing angles, as
given in Table \ref{convbestfit}, actually even with the expected
accuracy \cite{douglas}.
\begin{table}[!t]
\caption{Best fit to quark-lepton mass spectrum. All masses are
running masses at $1~\GeV$ except the top quark mass which is the
pole mass.}
\begin{displaymath}
\begin{array}{|c|c|c|}
\hline\hline
 & {\rm Fitted} & {\rm Experimental} \\ \hline
m_u & 4.4~\MeV & 4~\MeV \\
m_d & 4.3~\MeV & 9~\MeV \\
m_e & 1.6~\MeV & 0.5~\MeV \\
m_c & 0.64~\GeV & 1.4~\GeV \\
m_s & 295~\MeV & 200~\MeV \\
m_{\mu} & 111~\MeV & 105~\MeV \\
M_t & 202~\GeV & 180~\GeV \\
m_b & 5.7~\GeV & 6.3~\GeV \\
m_{\tau} & 1.46~\GeV & 1.78~\GeV \\
V_{us} & 0.11 & 0.22 \\
V_{cb} & 0.026 & 0.041 \\
V_{ub} & 0.0027 & 0.0035 \\ \hline
\Delta m^2_{\odot} & 9.0 \times 10^{-5}~\eV^2 &  5.0 \times 10^{-5}~\eV^2 \\
\Delta m^2_{\rm atm} & 1.7 \times 10^{-3}~\eV^2 &  2.5 \times 10^{-3}~\eV^2\\
\tan^2\theta_{\odot} &0.26 & 0.34\\
\tan^2\theta_{\rm atm}& 0.65 & 1.0\\
\tan^2\theta_{\rm chooz}  & 2.9 \times 10^{-2} & \sleq~2.6 \times 10^{-2}\\
\hline\hline
\end{array}
\end{displaymath}
\label{convbestfit}
\end{table}

\section{Conclusion}
\label{conclusion}

The hierarchical structure of the quark-lepton spectrum was
emphasized and interpreted as due to the existence of a mass
protection mechanism, controlled by approximately conserved chiral
flavour quantum numbers beyond the SM. The family replicated gauge
group model assigns a unique set of anomaly free gauge charges to
the quarks and leptons. With an appropriate choice of quantum
numbers for the Higgs fields, these chiral charges naturally
generate a realistic set of quark-lepton masses and mixing angles.
The top quark dominates the fermion mass matrices and we showed
how the Multiple Point Principle can be used to predict the top
quark and SM Higgs boson masses. We also discussed the lightest
family mass generation model, which gives simple and compact
formulae for all the CKM mixing angles in terms of the quark
masses.

\section*{Acknowledgements}
I should like to thank my collaborators Jon Chkareuli, Holger Bech
Nielsen and Yasutaka Takanishi for many discussions.

\end{document}